\documentclass[polish,english]{article}
\usepackage[T1]{fontenc}
\usepackage[latin2,latin9]{inputenc}
\usepackage{geometry}
\usepackage{float}
\usepackage{amsmath}
\usepackage{graphicx}
\usepackage{setspace}
\usepackage{babel}
\begin{document}
\noindent \begin{center}
{\LARGE Synchronous motion of two vertically excited planar elastic
pendula}
\par\end{center}{\LARGE \par}

\vspace{0.5cm}

\noindent \begin{center}
M. Kapitaniak$^{a,b*}$, P. Perlikowski$^{a}$, T. Kapitaniak$^{a}$
\par\end{center}

\noindent \begin{center}
\textit{\scriptsize $*$corresponding author: m.kapitaniak@abdn.ac.uk}
\par\end{center}{\scriptsize \par}

\vspace{0.2cm}

\begin{spacing}{0.20000000000000001}
{\scriptsize $^{a}$Division of Dynamics, Lodz University of Technology, Stefanowskiego 1/15, 90-924 Lodz, Poland}{\scriptsize \par}

{\scriptsize $^{b}$Centre for Applied Dynamics Research, School of
Engineering, University of Aberdeen, AB24 3UE, Aberdeen, Scotland,
United Kingdom}{\scriptsize \par}
\end{spacing}

\vspace{1cm}

\noindent \rule[0.5ex]{1\columnwidth}{0.5pt}

\subsection*{Abstract}

The dynamics of two planar elastic pendula mounted on the horizontally
excited platform have been studied. We give evidence that the pendula
can exhibit synchronous oscillatory and rotation motion and show that
stable in-phase and anti-phase synchronous states always co-exist. The complete bifurcational scenario leading from synchronous to asynchronous motion is shown.
We argue that our results are robust as they exist in the wide range
of the system parameters.

\noindent \textit{Keywords:} coupled oscillators, elastic pendulum,
synchronization

\vspace{0.2cm}

\noindent \rule[0.5ex]{1\columnwidth}{0.5pt}

\vspace{0.5cm}

\section{Introduction}

The elastic pendulum is a simple mechanical system which comprises
heavy mass suspended from a fixed point by a light spring which can
stretch but not bend when moving in the gravitational field. The state
of the system is given by three (spherical elastic pendulum) or two
(planar elastic pendulum) coordinates of the mass, i.e. the system
has three (spherical case) or two (planar case) degrees of freedom.
The equations of motion are easy to write but, in general, impossible
to solve analytically, even in the Hamiltonian case. The elastic pendulum
exhibits a wide and surprising range of highly complex dynamic phenomena
\cite{Anicin1993,Breitenberger1981,Carretero-Gonzalez1994,
Cayton1977,Cuerno1992,Davidovic1996,Holm2002,Kuznetsov1999,Lai1984,Lynch2002,
Lynch2002a,Lynch2004,Nunez-Yepez1990,Olsson1976,Pokorny2008,Rusbridge1980}. 

For small amplitudes perturbation techniques can be applied, the system
is integrable and approximate analytical solutions can be found. The
first known study of the elastic pendulum was made by Vitt and Gorelik
\cite{Vitt1933}. They considered small oscillations of the planar
pendulum and identified the linear normal modes of two distinct types,
vertical or springing oscillations in which the elasticity is the
restoring force and quasi-horizontal swinging oscillations in which
the system acts like a pendulum. When the frequency of the springing
and swinging modes are in the ratio $2:1$, an interesting non-linear
phenomenon occurs, in which the energy is transferred periodically
back and forth between the springing and swinging motions \cite{Anicin1993,Breitenberger1981,Carretero-Gonzalez1994,Cayton1977,Cuerno1992,Davidovic1996}.
The most detailed treatment of small amplitude oscillations of both
plane and spherical elastic pendula is presented in the works of Lynch
and his collaborators \cite{Holm2002,Lynch2002,Lynch2002a,Lynch2004}.
For large finite amplitudes the system exhibits different dynamical
bifurcations and can show chaotic behavior \cite{Kuznetsov1999,Lai1984,Nunez-Yepez1990,Olsson1976,Pokorny2008,Rusbridge1980}.

The dynamics of elastic pendulum attached to linear forced oscillator has been studied by Sado \cite{sado2004}. She has shown a one parameter bifurcation diagrams showing different behaviour of the systems (periodic, quasiperiodic and chaotic). According to our knowledge this is the only study of considered systems,   but one can find a
lot of papers concerning dynamics of classical pendulum attached to
linear or non-linear oscillator. Hatwal et al. \cite{Hatwal1982153,hatwal:657,hatwal:663}
gives approximate solutions in the primary parametric instability
zone, which allows calculation of the separate regions of periodic solutions.
Further analysis enables us to understand the dynamics around primary
and secondary resonances \cite{Bajaj1994,Cartmell1994173,Balthazar20011075,kecik2005,Song2003747}.
Then the analysis was extended to systems with non-linear base where
non-linearity is usually introduced by changing the linear spring
into nonlinear one \cite{Warminski2009612,WARMINSKI2001363,4723450,BVG2008}
or magnetorheological damper \cite{ISI:000289102700001}. Recently
the complete bifurcation diagram of oscillating and rotating
solutions has been presented \cite{Brzeski2012}. Dynamics of two coupled single-well Duffing oscillators forced by the common signal has been investigated in our previous papers  \cite{Perlikowski2008,Perlikowski2008c}. We have shown the
 detailed analysis of synchronization phenomena and compare different
methods of synchronization detection. 

In this paper we study the dynamics of two planar elastic pendula
mounted on the horizontally excited platform. Our aim is to identify
the possible synchronous states of two pendula. We give evidence that
the pendula can synchronize both in the oscillatory and rotational motion
moreover in-phase and anti-phase synchronizations co-exist. Our calculations
have been performed using software Auto--07p \cite{Doedel2011} developed
for numerical continuation of the periodic solutions and verified
by the direct integration of the equations of motion. We argue that
our results are robust as they exist in the wide range of the system
parameters. 

The paper is organized as follows. Sec. 2 describes the considered
model. We derive the equations of motion and identify the possible
synchronization states. In Sec. 3 we study the stability of different
types of synchronous motion. Finally Sec. 4 summarizes our results.

\section{Model of the system}

\begin{figure}
\begin{centering}
\includegraphics{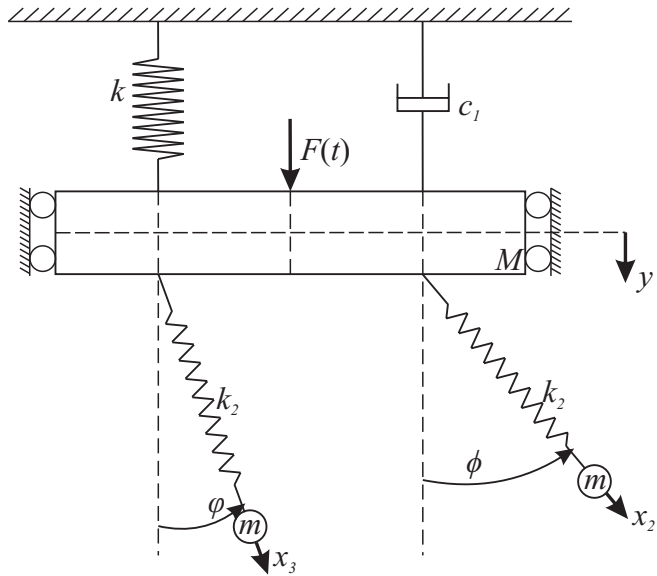}
\par\end{centering}

\caption{\label{fig:Model-of-the}Model of the system}
\end{figure}

The analyzed system is shown in Fig. \ref{fig:Model-of-the}. It consists
of two identical elastic pendula of length $l_{0}$, spring stiffness
$k_{2}$ and masses $m$, which are suspended on the oscillator. The
oscillator consists of a bar, suspended on linear spring with stiffness
$k_{1}$ and linear viscous damper with damping coefficient $c_{1}$.
The system has five degrees of freedom. Mass $M$ is constrained to
move only in vertical direction and thus is described by the coordinate
$y$. The motion of the first pendulum is described by angular displacement
$\varphi$ and its mass by coordinate $x_{2}$, that represent the
elongation of the elastic pendulum. Similarly the second pendulum
is described by angular displacement $\phi$ and its mass by coordinate
$x_{3}$. Both pendula are damped by torques with identical damping
coefficient $c_{2}$, that depend on their angular velocities (not
shown in Fig. \ref{fig:Model-of-the}). The small damping, with damping
coefficient $c_{3}$ is also taken for pendula masses. The system
is forced parametrically by vertically applied force $F(t)=F_{0}\cos\nu t$,
acting on the bar of mass $M$, that connects the pendula. Force $F_{0}$
denotes the amplitude of excitation and $\nu$ the excitation frequency. 

The equations of motion can be derived using Lagrange equations of
the second type. The kinetic energy $T$, potential energy $V$ and
Rayleigh dissipation $D$ are given respectively by:

\begin{spacing}{0.5}
\begin{flalign}
\hspace{1cm}\begin{array}{c}
T=\frac{1}{2}(M+2m)\dot{y}{}^{2}+\frac{1}{2}m\dot{x}_{3}^{2}+\frac{1}{2}m(l_{0}+y_{wst2}+x_{3})^{2}\dot{\phi}^{2}+m\dot{y}\dot{x}_{3}\cos\phi-m\dot{y}\dot{\phi}(l_{0}+y_{wst2}+x_{3})\sin\phi+\end{array} &  & {}
\end{flalign}

\begin{flalign*}
\hspace{1cm}+\frac{1}{2}m\dot{x}_{2}^{2}+\frac{1}{2}m(l_{0}+y_{wst2}+x_{2})^{2}\dot{\varphi}^{2}+m\dot{y}\dot{x}_{2}\cos\varphi-m\dot{y}\dot{\varphi}(l_{0}+y_{wst2}+x_{2})\sin\varphi &  & {}
\end{flalign*}

\end{spacing}

\vspace{0.5cm}

\begin{spacing}{0.5}
\begin{flalign}
\hspace{1cm}\begin{array}{c}
V=-mg(l_{0}+y_{wst2}+x_{2})\cos\varphi-mg(l_{0}+y_{wst2}+x_{3})\cos\phi+mg(l_{0}+y_{wst2})+mg(l_{0}+y_{wst2})+\end{array} &  & {}
\end{flalign}

\begin{flalign*}
\hspace{1cm}+\frac{1}{2}k_{1}(y+y_{wst1})^{2}+\frac{1}{2}k_{2}(y_{wst2}+x_{2})^{2}+\frac{1}{2}k_{2}(y_{wst2}+x_{3})^{2}-(M+2m)gy &  & {}
\end{flalign*}

\end{spacing}

\vspace{0.5cm}

\begin{spacing}{0.5}
\begin{flalign}
\hspace{1cm}D & =\frac{1}{2}C_{2}\dot{\varphi}^{2}+\frac{1}{2}C_{2}\dot{\phi}^{2}+\frac{1}{2}C_{3}\dot{x}_{2}^{2}+\frac{1}{2}C_{3}\dot{x}_{3}^{2} & {}
\end{flalign}

\end{spacing}

\vspace{0.5cm}

\noindent where $c_{3}$ is the damping coefficient of the pendulum
mass and $y_{wst1}=\frac{(M+2m)g}{k_{1}}$, $y_{wst2}=\frac{mg}{k_{2}}$
represent static deflation of mass $M$ and pendulums' mass $m$ respectively.
The system is described by five second order differential equations
given in the following form:

\begin{spacing}{0.5}
\begin{flalign}
\hspace{1cm}\begin{array}{c}
m(l_{0}+y_{wst2}+x_{2})^{2}\ddot{\varphi}+2m(l_{0}+y_{wst2}+x_{2})\dot{\varphi}\dot{x}_{2}-m\ddot{y}(l_{0}+y_{wst2}+x_{2})\sin\varphi+\end{array} &  & {}
\end{flalign}

\begin{flalign*}
\hspace{1cm}+mg(l_{0}+y_{wst2}+x_{2})\sin\varphi+C_{2}\dot{\varphi} & =0 & {}
\end{flalign*}

\end{spacing}

\vspace{0.5cm}

\begin{spacing}{0.5}
\begin{flalign}
\begin{array}{c}
\hspace{1cm}m(l_{0}+y_{wst2}+x_{3})^{2}\ddot{\phi}+2m(l_{0}+y_{wst2}+x_{3})\dot{\phi}\dot{x}_{3}-m\ddot{y}(l_{0}+y_{wst2}+x_{3})\sin\phi+\end{array} &  & {}
\end{flalign}

\begin{flalign*}
\hspace{1cm}+mg(l_{0}+y_{wst2}+x_{3})\sin\phi+C_{2}\dot{\phi} & =0 & {}
\end{flalign*}

\end{spacing}

\vspace{0.5cm}

\begin{spacing}{0.5}
\begin{flalign}
\hspace{1cm}m\ddot{x}_{3}+m\ddot{y}\cos\phi-m\dot{\phi}^{2}(l_{0}+y_{wst2}+x_{3})-mg\cos\phi+k_{2}(y_{wst2}+x_{3})+C_{3}\dot{x}_{3} & =0 & {}
\end{flalign}

\end{spacing}

\vspace{0.5cm}

\begin{spacing}{0.5}
\begin{flalign}
\hspace{1cm}m\ddot{x}_{2}+m\ddot{y}\cos\varphi-m\dot{\varphi}^{2}(l_{0}+y_{wst2}+x_{2})-mg\cos\varphi+k_{2}(y_{wst2}+x_{2})+C_{3}\dot{x}_{2} & =0 & {}
\end{flalign}

\end{spacing}

\vspace{0.5cm}

\begin{spacing}{0.5}
\begin{flalign}
\begin{array}{c}
\hspace{1cm}(M+2m)\ddot{y}+m\ddot{x}_{3}\cos\phi-2m\dot{x}_{3}\dot{\phi}\sin\phi-m(l_{0}+y_{wst2}+x_{3})\ddot{\phi}\sin\phi-m(l_{0}+y_{wst2}+x_{3})\dot{\phi}^{2}\cos\phi+\end{array} &  & {}
\end{flalign}

\begin{flalign*}
\hspace{1cm}+m\ddot{x}_{2}\cos\varphi-2m\dot{x}_{2}\dot{\varphi}\sin\varphi-m(l_{0}+y_{wst2}+x_{2})\ddot{\varphi}\sin\varphi-m(l_{0}+y_{wst2}+x_{2})\dot{\varphi}^{2}\cos\varphi+ &  & {}
\end{flalign*}

\begin{flalign*}
\hspace{1cm}-(M+2m)g+k_{1}(y+y_{wst1})+C_{1}\dot{y}-F_{0}\cos\nu t & =0 & {}
\end{flalign*}

\end{spacing}

\vspace{0.5cm}

\noindent In the numerical calculations we use the following values
of parameters: $M=10\hspace{0.05in}[\textrm{kg]}$, $m=0.2\hspace{0.05in}[\textrm{kg]}$,
$l_{0}=0.24849\hspace{0.05in}[\textrm{m]}$, $k_{1}=1642.0\hspace{0.05in}[\textrm{\ensuremath{\frac{N}{m}}]}$,
$k_{2}=19.7\hspace{0.05in}[\textrm{\ensuremath{\frac{N}{m}}]}$, $c_{1}=13.1\hspace{0.05in}[\textrm{\ensuremath{\frac{Ns}{m}}]}$
, $c_{2}=0.00776\hspace{0.05in}[\textrm{Nms]}$, $c_{3}=0.49\hspace{0.05in}[\textrm{\ensuremath{\frac{Ns}{m}}]}$,
$y_{wst1}=0.062\hspace{0.05in}[\mathrm{m}]$, $y_{wst2}=0.1\hspace{0.05in}[\mathrm{m}]$.

Introducing dimensionless time $\tau=\omega_{1}t$, where $\omega_{1}^{2}=\frac{k_{1}}{M+2m}$
is the natural frequency of mass $M$ with the attached pendula,
we obtain dimensionless equations of motion written as:

\begin{spacing}{0.5}
\begin{flalign}
\hspace{0.7cm}\ddot{\Psi}+\frac{2\beta_{2}}{(1+y_{2st}+\chi_{2})}\dot{\Psi}\dot{\chi}_{2}-\frac{\beta_{1}^{2}}{(1+y_{2st}+\chi_{2})}\ddot{\gamma}\sin\Psi+\frac{\sin\Psi}{(1+y_{2st}+\chi_{2})}+\frac{\alpha_{2}}{(1+y_{2st}+\chi_{2})^{2}}\dot{\Psi} & =0 & {}\label{eq:eq1}
\end{flalign}

\end{spacing}

\vspace{0.5cm}

\begin{spacing}{0.5}
\begin{flalign}
\hspace{0.7cm}\ddot{\Phi}+\frac{2\beta_{2}}{(1+y_{2st}+\chi_{3})}\dot{\Phi}\dot{\chi}_{3}-\frac{\beta_{1}^{2}}{(1+y_{2st}+\chi_{3})}\ddot{\gamma}\sin\Phi+\frac{\sin\Phi}{(1+y_{2st}+\chi_{3})}+\frac{\alpha_{2}}{(1+y_{2st}+\chi_{3})^{2}}\dot{\Phi} & =0 & {}\label{eq:eq2}
\end{flalign}

\end{spacing}

\vspace{0.5cm}

\begin{spacing}{0.5}
\begin{flalign}
\hspace{0.7cm}\ddot{\chi}_{3}+\frac{\beta_{1}^{2}}{\beta_{2}^{2}}\ddot{\gamma}\cos\Phi-\frac{1+y_{2st}+\chi_{3}}{\beta_{2}^{2}}\dot{\Phi}^{2}-\frac{1}{\beta_{2}^{2}}\cos\Phi+y_{st2}+\chi_{3}+\alpha_{3}\dot{\chi}_{3} & =0 & {}\label{eq:eq3}
\end{flalign}

\end{spacing}

\vspace{0.5cm}

\begin{spacing}{0.5}
\begin{flalign}
\hspace{0.7cm}\ddot{\chi}_{2}+\frac{\beta_{1}^{2}}{\beta_{2}^{2}}\ddot{\gamma}\cos\Psi-\frac{1+y_{2st}+\chi_{2}}{\beta_{2}^{2}}\dot{\Psi}^{2}-\frac{1}{\beta_{2}^{2}}\cos\Psi+y_{st2}+\chi_{2}+\alpha_{3}\dot{\chi}_{2} & =0 & {}\label{eq:eq4}
\end{flalign}

\end{spacing}

\vspace{0.5cm}

\begin{spacing}{0.5}
\begin{flalign}
\hspace{0.7cm}\begin{array}{c}
\ddot{\gamma}+\frac{\beta_{2}^{2}a}{\beta_{1}^{2}}\ddot{\chi}_{3}\cos\Phi-\frac{2\beta_{2}a}{\beta_{1}^{2}}\dot{\chi}_{3}\dot{\Phi}\sin\Phi-\frac{(1+y_{2st}+\chi_{3})a}{\beta_{1}^{2}}\ddot{\Phi}\sin\Phi-\frac{(1+y_{2st}+\chi_{3})a}{\beta_{1}^{2}}\dot{\Phi}^{2}\cos\Phi+\frac{\beta_{2}^{2}a}{\beta_{1}^{2}}\ddot{\chi}_{2}\cos\Psi+\end{array} &  & {}\label{eq:eq5}
\end{flalign}

\begin{flalign*}
\hspace{0.7cm}-\frac{2\beta_{2}a}{\beta_{1}^{2}}\dot{\chi}_{2}\dot{\Psi}\sin\Psi-\frac{(1+y_{2st}+\chi_{2})a}{\beta_{1}^{2}}\ddot{\Psi}\sin\Psi-\frac{(1+y_{2st}+\chi_{2})a}{\beta_{1}^{2}}\dot{\Psi}^{2}\cos\Psi-\frac{1}{\beta_{1}^{2}}+\gamma+y_{1st}+\alpha_{1}\dot{\gamma}-q\cos\mu\tau & =0 & {}
\end{flalign*}

\end{spacing}

\vspace{0.5cm}

\noindent where $\omega_{2}^{2}=\frac{k_{2}}{m}$, $\omega_{4}^{2}=\frac{g}{l_{0}}$,
$\mu=\frac{\nu}{\omega_{1}}$, $\beta_{1}=\frac{\omega_{1}}{\omega_{4}}$,
$\beta_{2}=\frac{\omega_{2}}{\omega_{4}}$, $a=\frac{m}{M+2m}$, $q=\frac{F_{0}}{\omega_{1}^{2}l_{0}(M+2m)}$,
$\alpha_{1}=\frac{C_{1}}{\omega_{1}(M+2m)}$, $\alpha_{2}=\frac{C_{2}}{m\omega_{4}l_{0}^{2}}$,
$\alpha_{3}=\frac{C_{3}}{ml_{0}\omega_{2}^{2}}$, $y_{1st}=\frac{y_{wst1}}{l_{0}}$,
$y_{2st}=\frac{y_{wst2}}{l_{0}}$, $\gamma=\frac{y}{l_{0}}$, $\dot{\gamma}=\frac{\dot{y}}{l_{0}\omega_{4}}$,
$\ddot{\gamma}=\frac{\ddot{y}}{l_{0}\omega_{4}^{2}}$, $\chi_{3}=\frac{x_{3}}{l_{0}}$,
$\dot{\chi}_{3}=\frac{\dot{x}_{3}}{l_{0}\omega_{2}}$, $\ddot{\chi}_{3}=\frac{\ddot{x}_{3}}{l_{0}\omega_{2}^{2}}$,$\chi_{2}=\frac{x_{2}}{l_{0}}$,
$\dot{\chi}_{2}=\frac{\dot{x}_{2}}{l_{0}\omega_{2}}$, $\ddot{\chi}_{2}=\frac{\ddot{x}_{2}}{l_{0}\omega_{2}^{2}}$,
$\Psi=\varphi$, $\dot{\Psi}=\frac{\dot{\varphi}}{\omega_{4}}$, $\ddot{\Psi}=\frac{\ddot{\varphi}}{\omega_{4}^{2}}$,
$\Phi=\phi$, $\dot{\Phi}=\frac{\dot{\phi}}{\omega_{4}}$, $\ddot{\Phi}=\frac{\ddot{\phi}}{\omega_{4}^{2}}$

The dimensionless parameters of the system have the following values:
$\beta_{1}=2$, $\beta_{2}=1.58$, $\alpha_{1}=0.1$, $\alpha_{2}=0.01$,
$\alpha_{3}=0.1$, $a=0.0192$, $y_{1st}=0.25$, $y_{2st}=0.4$.

We study system (9-13) in order to detect possible synchronization
ranges. There are two basic types of synchronous motion, which are
depicted in Fig. 2(a,b). The pendula can synchronize either in-phase
or in anti-phase with each other, i.e., $\Psi=\Theta$ or $\Psi=-\Theta$.
In both mentioned cases the forces acting in vertical direction on
mass $M$ are identical (there are no forces in horizontal direction),
hence the energy transmitted between mass $M$ and pendula in in-phase
and anti-phase motion is also identical. If there is an in-phase synchronization,
the anti-phase also coexists in the same range of parameters. The
accessibility of in-phase and anti-phase motion is governed only by
initial conditions. The pendula's masses are always synchronized in
the in-phase with each other, i.e., $\chi_{2}=\chi_{3}$. The anti-phase
configuration of the masses is not observed $(\chi_{2}=-\chi_{3})$ with
the oscillating pendula. The anti-phase synchronization of masses
is possible when the pendula are in equilibrium positions, then the
sum of forces transmitted to mass $M$ is equal to zero. 

\begin{figure}[H]
\begin{centering}
\includegraphics{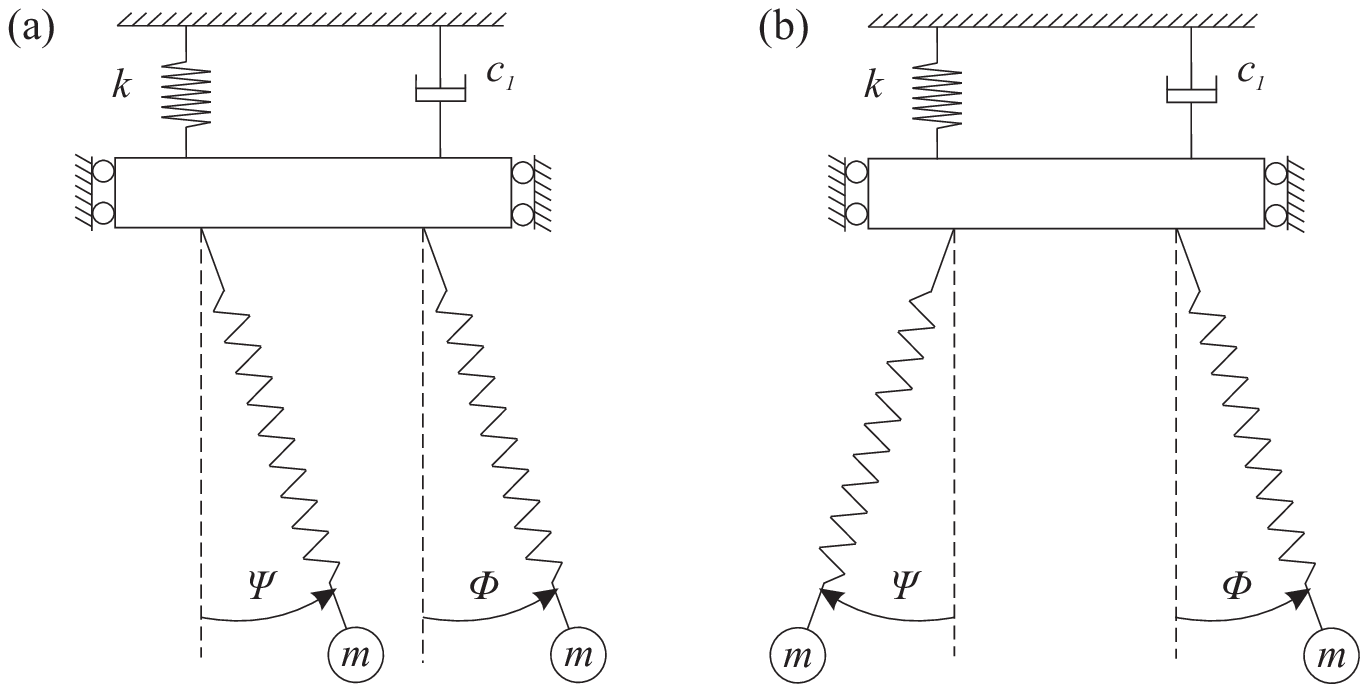}
\par\end{centering}

\caption{Possible synchronization (a) in-phase, (b) in anti-phase}
\end{figure}

\section{Stability of synchronous motion}

\subsection{Synchronous solutions in two dimensional parameters space}
\begin{figure}[H]
\begin{centering}
\includegraphics{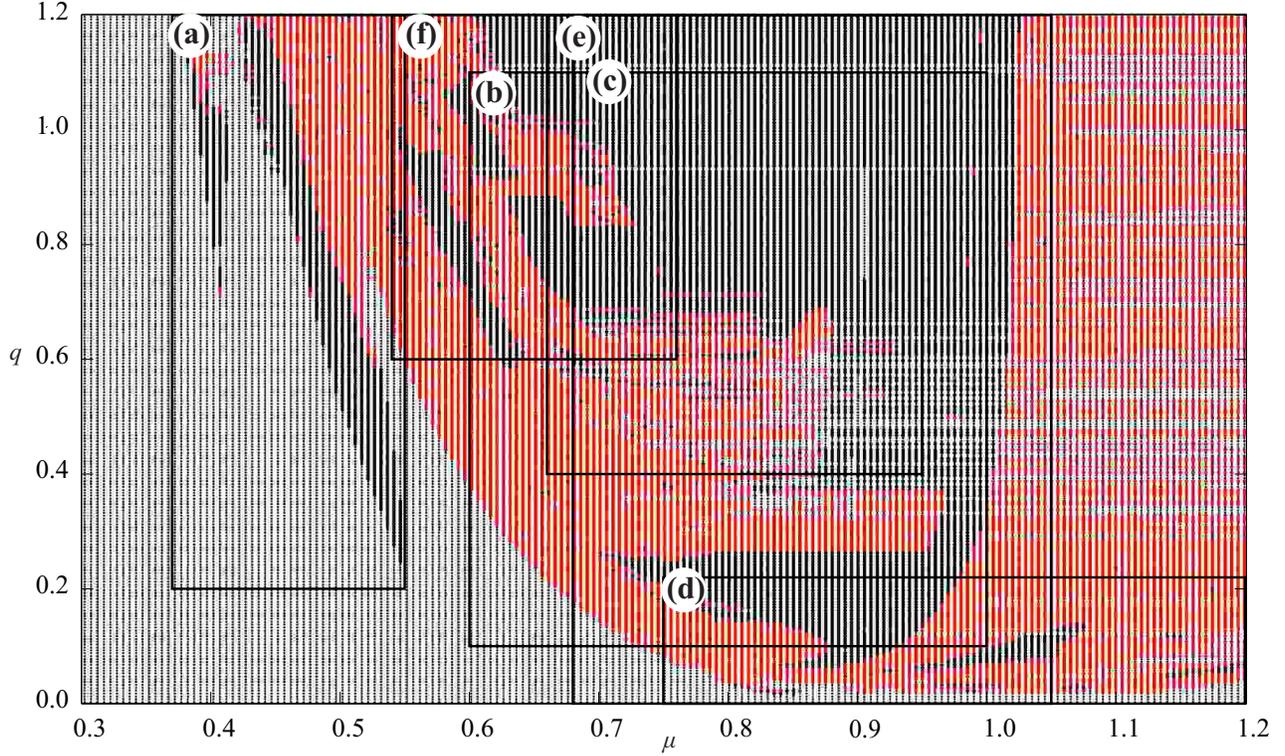}
\par\end{centering}

\caption{\label{fig:runge}(color online) The synchronous (black dots), asynchronous
(red dots) and equilibrium (small grey crosses) solutions of system
(9-13) in two parameters space: $\mu$ frequency and $q$ amplitude
of excitation. We calculate this plot by direct integration using
4th order Runge-Kutta algorithm. In rectangles (a-f) we highlighted
regions of synchronous motion calculated in Auto-07p (see Fig. \ref{fig:auto}).}
\end{figure}

\begin{figure}[p]
\begin{centering}
\includegraphics{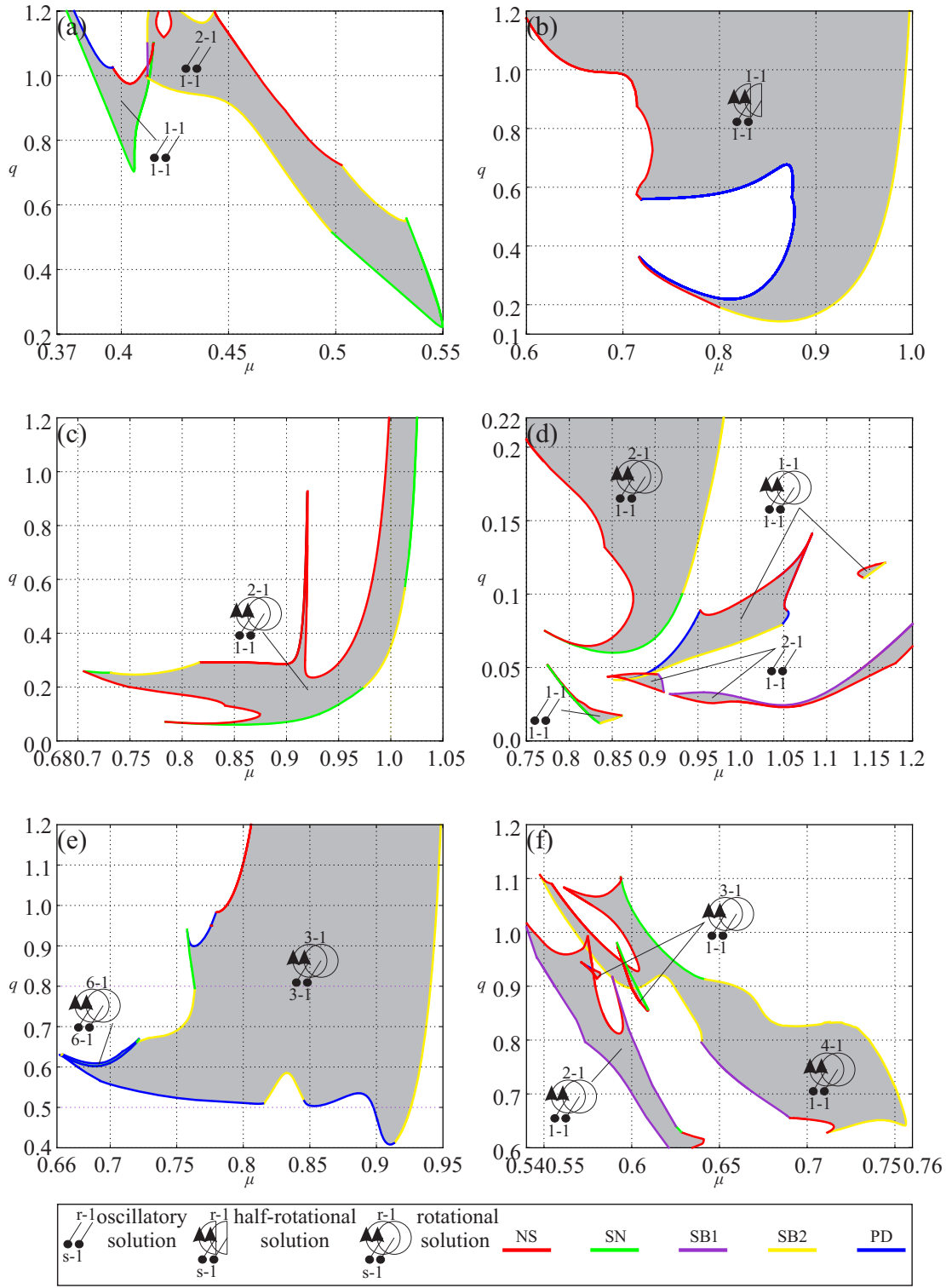}
\end{centering}

\caption{\label{fig:auto}(color) Stable ranges of synchronous motion calculated
in Auto-07p (see rectangles in Fig. \ref{fig:runge}). Color of lines
stand for different types of bifurcation: Neimark-Sacker (red), saddle-node
(green), pitchfork SB1 (violet), pitchfork SB2 (yellow) and period
doubling (blue). In region inside lines synchronous solutions are
periodic and stable. }
\end{figure}

In this section we study the stability of the observed synchronous
oscillations and rotations of the pendula. We present the bifurcation
diagrams calculated in two-parameter space: amplitude $q$ versus
frequency $\mu$ of excitation. We focus our attention on determining
the regions of synchronous stable motion and bifurcations that lead
to its destabilization. We consider the state of the system in the
following range $q\in[0.0,\:1.2]$ of forcing amplitudes and frequency
of excitation belonging to the range $\mu\in[0.3,\:1.2]$, which cover
the possible resonances in the system. Resonance should be observed
when the frequency of excitation comes close to the natural frequencies:
of mass $M$ equal to $\mu_{M}=1$, of pendula $\mu_{p}=0.50$ and
pendulum mass $\mu_{pm}=0.79$. We describe synchronous solutions
with respect to the forcing period according to ratio $r:1-s:1$, where
$r$ and $s$ denotes number of forcing periods, for which pendula
and pendula masses perform period one motion. Fig. \ref{fig:runge} presents
two parameter bifurcation diagram, obtained by direct integration of (9-13).
It shows the existence of synchronous, asynchronous motion and equilibrium
solutions. As soon as we have a lot of coexisting solutions to hold
clearance of Fig. \ref{fig:runge} we do not distinguish which type
of synchronous or asynchronous we find. By synchronous solution we
mean, that both pendula are in complete synchronization state, i.e.,
their amplitudes and frequencies are identical. For low amplitudes
of excitation, the only solution is equilibrium, which turns into
synchronous or asynchronous solution as the frequency of excitation
increases. The detailed analysis of synchronous solutions (shaded in grey color) was performed using continuation software Auto--07p \cite{Doedel2011}. We calculate the stability borders of each identified case, i.e., the ranges inside
which the given motion is stable. The first periodic solution is observed
for frequency of excitation equal to $\mu=0.406$ and for amplitudes
of excitation above $q=0.709$. This periodic solution is shown in
Fig. \ref{fig:auto}(a) is identified as synchronous oscillations
of pendula and pendula masses locked $1:1-1:1$ with forcing. This solution
is destabilized by saddle-node (green line), period doubling (blue
line) and Neimark-Sacker (red line) bifurcations curves. The continuation
reveals that for small range of parameters, around the frequency of
excitation close to the natural frequency of pendula, this solution
coexists with synchronous $2:1-1:1$ oscillations. Synchronous oscillations
$2:1-1:1$ are destabilized by saddle-node bifurcation curve then
by Neimark-Sacker and pitchfork symmetry breaking (SB2) bifurcations.
In the investigated system we distinguish two different symmetry breaking
pitchfork bifurcations one of them (SB2) brokes symmetry between the
pendula, the second one (SB1) brokes the symmetry of each pendula
but their motion remains identical \cite{Miles:1988:RSB:51797.51803}.
As the frequency of excitation increases we observe either asynchronous
motion or equilibrium. With further increase of excitation frequency
we observe asynchronous behavior, which change into two small regions
of synchronous rotations $3:1-1:1$. We show it in Fig. \ref{fig:auto}(f)
and this area is bounded by Neimark-Sacker, period doubling and saddle-node
bifurcations. This solution coexists with synchronous $2:1-1:1$ rotations,
presented also in Fig. \ref{fig:auto}(f). The stability region for
this solution is bounded by pitchfork SB1 bifurcation from the left
and right, Neimark-Sacker from above, and saddle-node and Neimark-Sacker
bifurcations from the right. Both these solutions coexist in small
range of considered parameters with another synchronous rotations
$4:1-1:1$, presented in Fig. \ref{fig:auto}(f). The synchronous
motion destabilizes from the right by saddle-node and pitchfork SB2
curves, from above by Neimark-Sacker, and from the left by Neimark-Sacker,
saddle-node and pitchfork SB2 curves. 

Around $\mu\approx 0.8$, where the resonance of pendulum masses occur,
the system possesses rich dynamics, which results in the coexistence
of different synchronous together with asynchronous solutions. This
includes synchronous rotations $2:1-1:1$ depicted in Fig. \ref{fig:auto}(c,d)
and synchronous $3:1-3:1$ rotations of pendula and pendula masses
presented in Fig. \ref{fig:auto}(e). The third which was found solution is synchronous half-rotations $1:1-1:1$ (Fig. \ref{fig:auto}(b)), for which both
pendula stop before approaching stable  and unstable equilibrium transferring
the whole energy into displacement of pendula masses. This multistability
causes that it is hard to compare the bifurcation diagrams from
the direct integration and Auto-07p. In the case
of rotations $2:1-1:1$ the synchronous motion is destabilzed from
the right by saddle-node and pitchfork SB2 curves, from above by Neimark-Sacker,
and from the left by saddle-node, Neimark-Sacker and pitchfork SB2
bifurcations. Synchronous rotations $1:1-1:1$ loose stability by
pitchfork SB2 from the right, and by Neimark-Sacker and period doubling
from the left. The synchronous rotations $3:1-3:1$ are mainly destabilized
by pitchfork SB2 from the right and by pitchfork SB2 and period doubling
from the bottom, and by period doubling, pitchfork SB2, saddle-node
and Neimark-Sacker from the left. From this solution, through period-doubling
bifurcation we find synchronized rotations $6:1-6:1$, shown in Fig.
\ref{fig:auto}(e). This solution is destabilized from above by period-doubling
bifurcation, from the left through pitchfork SB2 bifurcation, and
from below through saddle-node bifurcation (not visible, since coincides
with period-doubling boundary for rotations $3:1-3:1$).

As we pass through the resonance frequency of mass $M$ equal to $\mu=1$,
for higher amplitudes of excitation the only synchronous solution
includes $2:1-1:1$ synchronous rotations depicted in Fig. \ref{fig:auto}(c,d).
After the resonance, for amplitudes of excitation above $q=0.141$,
only asynchronous solutions are observed. Below this value, many small
synchronous regions were found. This includes synchronous $1:1-1:1$
oscillations and two regions of synchronous $2:1-1:1$ oscillations,
together with two regions of synchronous $2:1-1:1$ rotations. The
region of $1:1-1:1$ oscillations is enclosed by saddle-node, Neimark-Sacker
and pitchfork SB2 bifurcation curves. Oscillatory $2:1-1:1$ motion destabilizes
through pitchfork SB1 from above and Neimark-Sacker curves from below.
This solution coexists for small range of parameters with $2:1-1:1$ rotations,
which motion is destabilized by period doubling and Neimark-Sacker
from the left, and from the right by pitchfork SB2, Neimark-Sacker
and period doubling curves. We observe the excellent correlation in
these regions between the results from numerical continuation and
direct integration. 

\noindent 

\noindent 

\subsection{One parameter continuation}

\begin{figure}
\begin{centering}
\includegraphics{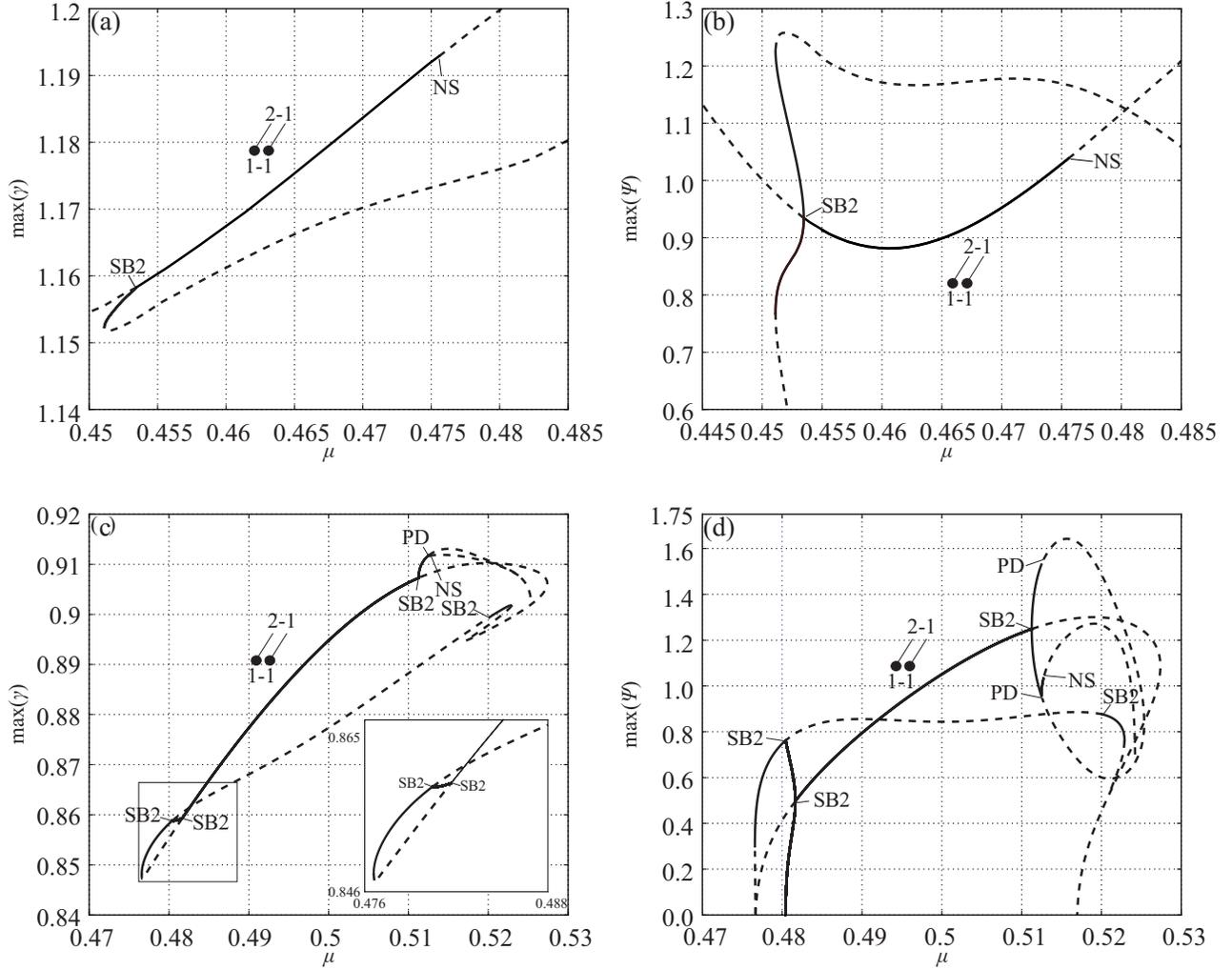}
\par\end{centering}

\caption{\label{fig:1-parameter-continuation2}1 parameter continuation of
fully synchronized: 2:1-1:1 oscillations ((a) mass $M$, (b) pendulum
1, \foreignlanguage{polish}{$q=0.899$, $\mu=0.455$}), 2:1-1:1 oscillations
((c) mass $M$, (d) pendulum 1, \foreignlanguage{polish}{$q=0.654$,
$\mu=0.5$}). The continuous and dashed lines correspond to stable
and unstable periodic solutions respectively. Abbreviations depicted
following bifurcations: NS (Neimark-Sacker), PD (period doubling),
SB1 (pitchfork SB1) and SB2 (pitchfork SB2). Other changes of the
stability take place through the saddle-node bifurcations.}
\end{figure}

\begin{figure}
\begin{centering}
\includegraphics{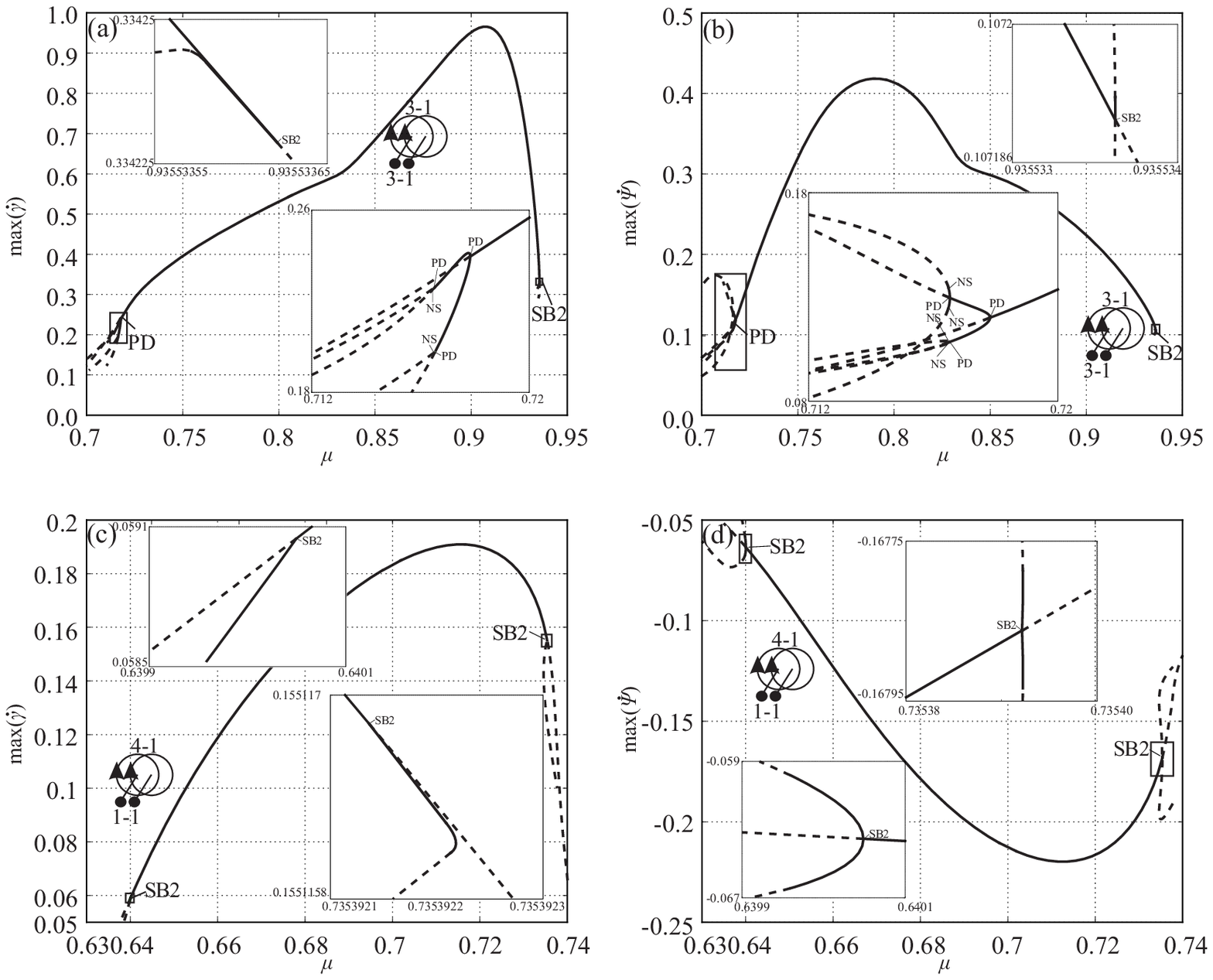}
\par\end{centering}

\caption{\label{fig:1-parameter-continuation}1 parameter continuation of fully
synchronized: $3:1-3:1$ rotations ((a) mass $M$, (b) pendulum 1,
\foreignlanguage{polish}{$q=0.654$, $\mu=0.8$}), $4:1-1:1$ rotations
((c) mass $M$, (e) pendulum 1,\foreignlanguage{polish}{ $q=0.8$,
$\mu=0.7$}). The continuous and dashed lines correspond to stable
and unstable periodic solutions respectively. Abbreviations depicted
following bifurcations: NS (Neimark-Sacker), PD (period doubling),
SB1 (pitchfork SB1) and SB2 (pitchfork SB2). Other changes of the
stability take place through the saddle-node bifurcations.}
\end{figure}

In this subsection we show one parameter continuation of four periodic
solutions (two oscillating and two rotational) for fixed amplitude
of excitation, as a bifurcation parameter we choose the frequency
of excitation $\mu$. We start each path-following on the periodic solution and continue in two directions (forward and backward). In Fig. \ref{fig:1-parameter-continuation2}(a-d)
we present the synchronized oscillating periodic solutions, their
regions of stability are shown in Fig. \ref{fig:auto} (a). System (\ref{eq:eq1}-\ref{eq:eq5}) is given by five second order ODEs,
hence the phase space is ten dimensional and at least five figures
(amplitude of each degree of freedom) are necessary to show its complete
dynamics. To decrease it we focus on the dynamics of first pendula (second pendula has the same amplitude in the synchronized state) and mass $M$.
The first presented branch of periodic solutions in Fig. \ref{fig:1-parameter-continuation2}(a,b)
is synchronous $2:1-1:1$ oscillations, in previous subsection we
show that this family is destabilized by Neimark-Sacker bifurcation
from the right and from the left by pitchfork symmetry braking SB2.
Switching the branch at pitchfork bifurcation enables us to find another
stable branch of asynchronized oscillations $2:1-1:1$, that
looses its stability through the saddle-node bifurcation. After pitchfork
symmetry breaking SB2 bifurcation the solutions of one pendulum is
located at upper branch (see Fig. \ref{fig:1-parameter-continuation2}(b))
and the second pendulum on lower branch or vice-versa. The branch synchronous
oscillations $2:1-1:1$, shown in Fig. \ref{fig:1-parameter-continuation2}
(c,d), present much richer scenario than other. These oscillations
destabilize from both sides through pitchfork SB2 bifurcation. When
we switch branch in left SB2 point, we find family of stable
asynchronized oscillations $2:1-1:1$. Finally, when the amplitudes of pendula
reach zero their motion stops. When we continue in the opposite direction the stability
is lost in pitchfork SB2 bifurcation. Another change of branch allows
us to observe another asynchronous periodic solutions, for which first pendulum
oscillates $2:1-1:1$, second pendulum is at rest (not shown here) and pendula
masses oscillate $1:1-1:1$ in asynchronized manner. One end of this stable
branch destabilizes through saddle-node bifurcation and the second
one by pitchfork SB2 bifurcation. As the frequency of excitation increases
the stability of this solutions is regained through pitchfork SB2 bifurcation
and lost again through saddle-node bifurcation. Note that for the
mass $M$, the bifurcation points that are responsible for the destabilization
of periodic solutions for synchronized oscillations $2:1-1:1$ and
asynchronized oscillations $2:1-1:1$, are placed very close to
each other. When we switch the branch in the right SB2 bifurcation
point of the synchronized oscillations $2:1-1:1$, we find asynchronized
solution of oscillations $2:1-1:1$ that persists for small interval
of excitation frequency. It destabilizes from above and below through
period-doubling bifurcations. Switching the branch in lower
period doubling bifurcation point enables us to observe asynchronized
oscillations $4:1-2:1$, that destabilize through Neimark-Sacker bifurcation.
The bifurcation diagram, shown in Fig. \ref{fig:1-parameter-continuation}(a-b),
shows $3:1-3:1$ rotational periodic solutions for $q=0.654$. In
contrary to the previous cases, to hold a physical meaning, on the
horizontal axis we plot the amplitude of velocity. The stability region
of this branch is shown in Fig. \ref{fig:auto}(e). This family of solutions
looses its stability through period-doubling and pitchfork SB2 bifurcations.
Switching the branch in the period-doubling bifurcation point, allows
to observe another period doubling, which leads us to two branches
of synchronized $6:1-6:1$ rotational solution. This branch
looses its stability once again through period-doubling bifurcation. After another switch of branch in period doubling bifurcation point, we
reach two synchronized periodic $12:1-12:1$ rotational
solutions. They are stable in very narrow range of excitation frequency
and loosing stability via Neimark-Sacker bifurcations. After switching
the branch in right pitchfork SB2 bifurcation point, we observe asynchronized
rotations $3:1-3:1$, that are stable in very small interval, finally
loosing its stability through saddle-node bifurcation. In Fig. \ref{fig:1-parameter-continuation}(c-d) we present synchronized
rotations $4:1-1:1$, that loose stability through pitchfork symmetry
braking from the right and left. Switching the branch in both SB2
bifurcation points let us to find asynchronous rotations $4:1-4:1$,
that are stable in very narrow interval of excitation frequency, loosing
finally stability through saddle-node bifurcation.

\section{Conclusions}

In the system of two planar elastic pendula suspended on the excited
linear oscillator one can observe both in-phase and anti-phase synchronization
of the elastic pendula. In-phase and anti-phase synchronous states always co-exist.
Pendula can synchronize during the oscillatory and rotational motion
but only when their behaviour is periodic. We have not observed the
synchronization of the chaotically behaving pendula. This result is
contrary to the great number of chaos synchronization examples \cite{Stefanski2003,Stefanski2000,Maistrenko1997}
but confirms the results obtained in \cite{Czolczynski2007} where
it has been shown that the forced Duffing\textquoteright{}s oscillators
mounted to the elastic beam can synchronize only after motion become
periodic. The synchronization of the chaotic motion of the pendula
is impossible as the excited oscillator transfers the same signal
to both pendula which cannot differently modify the pendula\textquoteright{}s
motion. We also have not observed in-phase or anti-phase synchronization
of the pendula when masses $m_{2}$ and $m_{3}$ are in anti-phase. With one parameter bifurcation diagrams we present a bifurcational scenario of synchronous solutions. We show a route from synchronous via asynchronous periodic solutions to quasiperiodic and chaotic behaviour.
In this case the pendula in-phase or anti-phase synchronization is
impossible as the pendula have different and non-constant lengths. In this case one
can expect some kind of generalized synchronization but this problem
will be addressed elsewhere \cite{Kapitaniak}.

We show two dimensional bifurcation diagrams with the most representative
periodic solutions in the considered system. In the neighbourhood of
the linear resonances of subsystems we have rich dynamics with both
periodic and chaotic attractors\cite{ChudzikPSK11}. Our results are
robust as they exit in the wide range of system parameters, especially
two dimensional bifurcation diagram can be used as a scheme of bifurcations
in the class of systems similar to investigated in this paper.

\section*{Acknowledgement}

This work has been supported by the Foundation for Polish Science,
Team Programme (Project No TEAM/2010/5/5).

\bibliographystyle{elsarticle-num}
\nocite{*}

\end{document}